# WHAT ARE THE HIDDEN QUANTUM PROCESSES BEHIND NEWTON'S LAWS?
EMQG, CA Theory, Quantum Inertia, and Newtonian Physics

**Tom Ostoma and Mike Trushyk**
Email: emqg@rogerswave.ca
Monday, April 19, 1999

## ABSTRACT

We investigate the hidden quantum processes that are responsible for Newton's laws of motion and Newton's universal law of gravity. We apply Electro-Magnetic Quantum Gravity or EMQG (ref. 1) to investigate Newtonian classical physics. EQMG is a quantum gravity theory that is manifestly compatible with Cellular Automata (CA) theory, a new paradigm for physical reality. EMQG is also based on a theory of inertia (ref. 5) proposed by R. Haisch, A. Rueda, and H. Puthoff, which we modified and called Quantum Inertia (QI). Quantum Inertia theory states that in Newton's $2^{nd}$ law of motion (F=MA), inertia is caused by the strictly local electrical force interactions of matter (ultimately composed of electrically charged quantum particles) with the surrounding electrically charged virtual particles of the quantum vacuum. When an electrically charged particle is accelerated, an electrical force results between the particle and the surrounding electrically charged virtual particles of the quantum vacuum appears in a direction to oppose the acceleration. The sum of all the tiny electrical forces originating between each charged particle and the surrounding quantum vacuum, is the source of the <u>total inertial force</u> of a mass which opposes accelerated motion in Newton's F = MA. Quantum Inertia theory resolves the problems and paradoxes of accelerated motion introduced in Mach's principle by suggesting that the state of acceleration of the charged virtual particles of the quantum vacuum with respect to a mass, serves the function of Newton's <u>*absolute space*</u> for accelerated masses only. Newton's three laws of motion result as follows:

- **Newton's First Law of Motion (Bodies at Rest / Bodies in Motion):** First, if a mass is at relative rest with respect to some observer in deep space, and if no external forces acts on the mass to overcome the opposing force from the quantum vacuum, the electrically charged elementary particles that make up the mass maintain a *net average acceleration* of zero with respect to the electrically charged virtual particles of the quantum vacuum (from frequent electrical scattering with the vacuum). This means that no change in velocity is possible, and the mass remains at rest. Secondly, if a mass has been given a constant velocity with respect to an observer in deep space, and again if no external forces act on the mass, the electrically charged elementary particles that make up the mass *still maintains a net <u>acceleration of zero</u>* with respect to the electrically charged virtual particles of the quantum vacuum. No change in velocity is possible without the presence of an external force to overcome the electrical resistance force.
- **Newton's Second Law (F=MA):** This law is a direct consequence of Quantum Inertia theory.
- **Newton's Third Law of Motion (Action-Reaction Forces):** According to the boson force particle exchange paradigm of quantum field theory all forces result from particle exchanges, including mechanical contact forces. Therefore the force that body 1 exerts on body 2 (called the action force), is the result of the emission of force exchange particles (usually photons) from the electrically charged particles that make up body 1, which are readily absorbed by the electrically charged particles that make up body 2, resulting in a force acting on body 2. Similarly, the force of body 2 on body 1 called the reaction force is the result of the absorption of force exchange particles that are originating from the electrically charged particles that make up body 2, and absorbed by the electrically charged particles that make up body 1, resulting in a force acting on body 1. An important property of electrical charge is the ability to readily emit <u>and</u> absorb photon particles. Therefore body 1 is both an emitter and also an absorber of the photon force exchange particles, and body 2 is also both an emitter and an absorber of the force exchange particles. This is the reason there is action and reaction forces in all mechanical force actions.

- **Newton's Universal Gravitational Law ($F = Gm_1m_2/r^2$):** The gravitational force 'F' originates from the exchange of gravitons between the two mass particles $m_1$ and $m_2$ (possessing 'mass charge') separated by a distance 'r', with G being Newton's universal gravitational constant. Each exchanged graviton imparts a tiny impulse of momentum to the absorbing mass particle, which after countless exchanges produces a smooth force. Each mass is <u>both</u> an emitter and a absorber of gravitons, because all mass particles possesses the property of 'mass-charge'. The flux of gravitons emitted by a given mass particle is a fixed quantity, that depends only on the magnitude of the charge, and is defined as the *'mass charge'* of the mass particle. This two way force interaction process causes the affected mass particles to accelerate towards each other. The strength of the gravitational force varies as the inverse square of the distance of separation between the two mass particles due to a geometric spreading of the graviton particles in the following way. Each mass particle sends and receives gravitons. As a source of gravitons, the mass particle emits gravitons in all directions. The number of gravitons per unit area absorbed at the destination particle distance r decreases by a factor $1/4\pi r^2$ (the surface area of a sphere, since gravitons spread equally in all directions) at a distance 'r'. If the distance doubles, the number of gravitons exchanged decreases by a factor of four, and this is why there is an inverse square law. If the mass $m_1$ (or $m_2$) of a particle is doubled, the 'mass charge' doubles and twice as many gravitons are exchanged. This accounts for the linear product of mass terms in the numerator of Newton's inverse square law. The constant G is the proportionality constant, and can be derived (in principle) if one knows the detailed characteristics of the graviton.

- **Equivalence of Newtonian Inertial mass and Gravitational mass:** The equivalence of inertial and gravitational mass, first proposed by Newton and later refined by Einstein, has remained a mystery until the arrival of EMQG. Mass equivalence ultimately results from the reversal of the relative acceleration vectors of the electrically charged mass particles that make up an accelerated mass, <u>with respect</u> to the (statistical) average acceleration of the electrically charged virtual particles of the quantum vacuum. This occurs when one changes from an accelerated to a gravitational frame. On the earth ($M_e$), the magnitude of the gravitational field acceleration is the same as the virtual mass particle acceleration of the quantum vacuum: namely $A=GM_e/r^2$, which is also the same acceleration chosen for a rocket accelerating a mass 'M'. From the reference frame of an average accelerated virtual mass particle falling on the earth, a virtual particle 'sees' the real particles that constitutes a stationary mass 'M' on the earth accelerating in exactly the same way as an average stationary virtual mass particle in the rocket 'sees' the accelerated mass particles constituting the mass 'M' on the floor of the rocket. In other words, from the perspective of a real particle on mass 'M', the <u>relative state of acceleration</u> of the virtual particles of the quantum vacuum appears the same in both reference frames, and hence we have equivalence of inertial and gravitational mass. Mathematically, this can be seen in Newton's laws by slightly rearranging Newton's formulas as follows:

$F_i = M_i (A_i)$ ... the inertial force $F_i$ opposes the acceleration $A_i$ of mass $M_i$ in the rocket, caused by the sum of the tiny electrical force interactions with the electrically charged virtual particles.

$F_g = M_g (A_g) = M_g (GM_e/r^2)$ ... the gravitational force $F_g$ is the result of <u>an equivalent inertial type force</u> given by $M_g A_g$, where $A_g = GM_e/r^2$ is now the <u>downward acceleration</u> of the virtual particles of the quantum vacuum, which are being accelerated by graviton exchanges with earth. Since $F_i=F_g$, and since the acceleration of the rocket is chosen to be the same as the acceleration of gravity on the earth; i.e. $A_g = A_i = GM_e/r^2$, therefore $M_i=M_g$, the mass equivalence principle.

## 1. INTRODUCTION

*'... and the Newtonian scheme was based on a set of assumptions, so few and so simple, developed through so clear and so enticing a line of mathematics that conservatives could scarcely find the heart and courage to fight it.'*

*- Isaac Asimov*

Newton's basic laws of motion are taught to all beginning students of physics. They have formed the central basis of classical physics for hundreds of years. Newton's laws of motion and Newton's universal gravitational law are so powerful that they are still used today to accurately guide spacecraft to their respective destinations. General relativity, or for that matter, a quantum theory of gravity is not really needed by NASA for any of their space exploration activities, or for any normal day to day activities. So then why should anyone want to investigate any possible quantum processes behind gravity or behind Newton's highly successful classical laws of motion? There are several reasons.

Newton's laws are basically about matter, and also about forces. Particularly they are about inertial forces and gravitational forces. Space is taken to be an absolute, passive backdrop, which is totally unaffected by the activities of matter, or of forces. Time is also taken to be absolute and immutable in the Newtonian scheme. However, a lot of developments have occurred in physics since Newton first proposed his basic laws. Physics has undergone a major revolution in the way we think about matter, forces, and also about space and time.

Since the early 1900's quantum theory has emerged as a very powerful tool to understand the behavior of matter on the smallest distance scales. It forms our ***first major revolution*** of the nature of physical reality. Not long after Newton proposed his famous laws it was realized that matter was composed of quantum particles that move in ways that would have been very unfamiliar to Newton. Quantum theory teaches us that, in general the world is not perfectly smooth and continuous. It shows us that matter is quantized in the form of elementary particles, and that energy too is quantized. Shortly after quantum theory, quantum field theory was developed which revolutionized the way we understand forces to work. It turns out that all forces are also quantized, and that discrete 'packets of force' called vector bosons are exchanged between particles. Each force particle transfers a tiny impulse of momentum to the destination particle. Force particles are exchanged in such huge numbers that a smooth force results, which we feel in many of our day to day activities. Although quantum theory, as is currently formulated, is based on an infinitesimal space-time continuum, which many believe should also be quantized.

The ***second major revolution*** of physical reality came with the introduction of special relativity by Albert Einstein. This completely transformed our way of thinking about space and time. By accepting that light is the ultimate speed, it turns out that space and time had to be united into a Minkowski, 4D space-time. More importantly, 4D space-time is no longer an absolute backdrop that Newton envisioned, but is changeable and is relative, and space-time changes value with different observers in different reference frames (or even under the action of a gravitational field). Shortly afterwards, Einstein developed general relativity in order to come up with a theory of gravity that was compatible with the Minkowski, 4D space-time of special relativity, and also to provide a generally covariant theory of gravity, where the laws of gravity are independent of the frame of reference or coordinate system chosen by the observer. This resulted in the adoption of the quasi-Riemannian geometry for gravity, in which 4D space-time is curved near the presence of a large mass.

Is there a common foundation that marries relativity and quantum theory into a common framework, while being able to explain how Newton's laws emerge from some sort of hidden quantum process in a satisfactory fashion? Currently it seems that the important core concepts in relativity and in quantum theory are not compatible, and a new theory seems to be required to merge these two very different views of physical reality.

ElectroMagnetic Quantum Gravity or EMQG (ref. 1) was developed to provide such a quantum model of gravity that we believe successfully marries relativity and quantum theory. The subject of this paper is to use EQMG to reveal the hidden quantum processes behind Newtonian physics. EMQG theory was also developed to be manifestly compatible with a Cellular Automaton model of the universe (ref. 2 and 4), which forms our **third, and new major revolution** of the nature of physical reality.

EMQG is based totally on the particle exchange paradigm of quantum field theory for *all* the forces of nature, where gravity is no exception. The particle exchange paradigm fits very well within the Cellular Automata framework. The notion that our universe is a vast Cellular Automaton started with the work of Edward Fredkin in the 1980's, who is the first person credited with introducing the idea that our universe is a huge cellular automaton computer simulation. Cellular Automaton (CA) is basically a computer model of the universe, where space consists of a large number of cells, which are the storage locations for numbers. The CA model automatically quantizes space, and also time, and thus completing the quest to quantize all key physical quantities.

In a CA computer simulation each and every cell contains some initial numbers, and the same set of mathematical rules that are applied for *each* and *__every__* cell. The rules are the mathematical operations that specify how these numbers are to be changed at the next computer '*clock*' interval. These logical rules or program contained in each cell specifies the new state of that cell on the next 'clock' period, which is based on the cell's current numerical state, and on that of all the cell's immediate neighbors (each cell in CA has a fixed number of neighboring cells which relates to the dimensionality of our space).

The CA model provides a model of physics that is capable of unifying all of physics. It has the potential to provide a single 'mechanistic' model of the universe. However, the Cellular Automaton mechanism would have been totally alien to Newton and his mechanical clock-work universe. We believe that when CA theory is developed fully, an exact CA model will be found that models everything precisely, and can used to explain all physical reality. EMQG theory was the result of intensive investigations into a model for gravity that fits the general principles of CA theory. EMQG is also based on the particle exchange paradigm, which also fits well with the CA concept. Let us briefly explore the physical consequences of living in a universe that is a vast cellular automaton simulation.

## 2. SPACE, TIME, MATTER AND CELLULAR AUTOMATA THEORY

*"By getting to smaller and smaller units, we do not come to fundamental units, or indivisible units, but we do come to a point where division has no meaning."*

*Werner Heisenberg*

*" It always bothers me that, according to the laws as we understand them today, it takes a computing machine an infinite number of logical operations to figure out what goes on in no matter how tiny a region of space and no matter how tiny a region of time. .... why should it take an infinite amount of logic to figure out what one tiny piece of space-time is going to do?"*

*- Richard Feynman*

Cellular Automata (CA) theory (references 2 and 4) forms the basis of EMQG, and has guided us towards our theory of quantum gravity and to a new understanding of Newton's laws of motion. Cellular Automata theory will be discussed fully in section 3. According to CA theory, **all** the laws of physics must be the result of interactions that are strictly local (changes in CA state propagate from cell through neighbor cell), which therefore forbids any kind of action at a distance. If CA theory is correct, then the global laws of physics (like the Newton's law of inertia 'F=MA') should be the result of the local actions of matter particles, which exist as 'information patterns' in the cells of the CA. In addition, CA theory suggests that space, time, matter, energy and motion are all the *same* thing, namely the result of information changing state according to some set of specific mathematical rules.

If CA theory is true, then this implies that the familiar ideas of space, time, and matter are not the basic elements of reality. In CA theory, all physical phenomena turn out to be the end result of a vast amount of numerical information that is being processed on a 'universal computer', and that the information *inside* this computer is in fact our *whole* universe, including ourselves! All elements of reality are due to numerical information being processed on the 'cells' at incredibly 'high speeds' (with respect to our perception of time). The computer hardware is forever inaccessible to us, because we ourselves are also information patterns residing in the cells. The laws of physics that govern the 'hardware' functioning of the universal computer hardware do not even have to be the same as our own physical laws. In fact, the computer 'hardware' that governs our reality can be considered by definition of the word 'universe' to be outside our own universe, and therefore *inaccessible.*

The computer model that best fits the workings of our universe is quite different from that of an ordinary personal desktop computer. In fact, our universe is implemented on the most massively parallel computer model currently known to computer science. This parallel computer model is called a 'CELLULAR AUTOMATON'. A CA consists of a huge array of 'cells' (or memory locations) that are capable of storing numerical data, which change state on every clock period everywhere, according to the rules or program. This type of computer was discovered theoretically by Konrad Zuse and Stanislav Ulam in the late 1940's, and later put to use by John von Neumann to model the real world behavior of complex spatially extended structures. The best known example of a CA is the game of life, which was originated by John Horton Conway.

In fact, a CA computer is so powerful that it is capable of updating it's entire memory (no matter how big) in a single clock cycle! Contrast this to the desktop computer, which takes millions of clock cycles to update the entire memory. A CA is inherently symmetrical, because one set of rules is programmed and repeated for each and every memory cell. We believe that this accounts for the high degree of spatial symmetry found in our universe. In other words, the laws of physics are the same no matter where you are, or how you are rotated in space. This is accountable by the perfect symmetry of the cell space. Not only is the CA a vastly parallel computer, but it is also the ***fastest*** known parallel computer processor, and therefore an ideal choice to 'build' a universe.

On remotely small scales of distance and time, called the Plank scale (about $10^{-35}$ meters distance and $10^{-43}$ seconds) the 'view' of the universe is completely unlike what we know to exist from our basic senses. Space, time, and matter no longer exist as separate entities. Elementary particles of matter reveal themselves as oscillating numeric information patterns. Motion turns out to be an illusion, which results from the 'shifting' of particle-like information patterns from cell to cell. However, there is no real motion in a CA! There is simply the transfer of information from place to place. Forces, which result from vector boson particle exchanges, reveal themselves as the exchange of oscillating information patterns, which are readily emitted and absorbed by oscillating matter particles. The absorption of a vector boson information pattern changes the internal oscillation of a particle, and causes an 'acceleration' to occur along a direction towards, or away from the source.

The quantization of space reveals itself in the cells, or storage locations for the numbers of the computer. What causes the numbers to change state as the regular clock intervals progresses? It is the same mathematical and local rules that are preprogrammed in all the storage cells of the computer. All the cells change state at the same time (not in our time, but during a CA clock transition time) and at regular CA 'clock' intervals (not to be confused with clocks in our universe). Thus in CA theory space at the lowest scales isn't nothing, it is something; it is vast numbers of memory cells. Particle numeric information patterns residing in the cells are dynamic and shifting, but they are ***not actually*** moving! They simply change state as the computer simulation evolves. In order to understand CA theory and its relationship to modern physics we will need to take a brief look at the subject of Cellular Automata theory.

## 3. CELLULAR AUTOMATA THEORY

*"Digital Mechanics is a discrete and deterministic modeling system which we propose to use (instead of differential equations) for modeling phenomena in physics.... We hypothesize that there will be found a single cellular automaton rule that models all of microscopic physics; and models it exactly. We call this field Digital Mechanics."*

<div align="right">

*Edward Fredkin*

</div>

*"All (the universe) is numbers"*

<div align="right">

*- Pythagoras*

</div>

In this section we provide a brief introduction to Cellular Automata theory, and it's connection with modern physics (references 2 and 4). We also include some new material we developed in regards to the CA space dimensionality. Edward Fredkin (ref. 34) has been the first person credited with the idea that our universe is a huge cellular automaton computer simulation. The cellular automaton (CA) model is a special computer that consists of a large number of cells, which are storage locations for numbers. Each cell contains some initial numerical state, and the same set of rules are applied for *each* and every cell. The rules specify how these numbers are to be changed at the next computer 'clock' interval. Mathematically, a 'clock' is required in order to synchronize the next state of all the cells. This 'clock' can be thought as being a marker to let a cell know when to respond to the state of the immediate surrounding cells. The logical rules of a cell specifies the new state of that cell on the next 'clock' period, based on the cells current state, and on that of all the cell's immediate neighbors (each cell in CA has a fixed number of neighboring cells). The number of neighbors that influence a given cell is what we call the basic 'connectivity' of the cellular automaton. The number of neighbors that connect (or influence) a given cell is defined as the CA connectivity. The connectivity can be any positive, non zero integer number.

We define a 'geometric' cellular automaton as a CA configuration where each cell has the correct number of neighbors such that the CA connectivity allows a simple cubic geometric arrangement of cells. This can be visualized as the stacking of cells into squares, cubes, hypercubes, etc. This structure is well suited for constructing a 3D space. The mathematical spatial dimensions required to contain the geometric CA is defined as the 'dimensionality of the space' for a geometric cellular automaton. For example, in a 2D geometric CA each cell has 8 surrounding neighbors, which can be thought of as forming a simple two dimensional space. One set of rules exists for every given cell, based on the input from its immediate 8 neighbors (and possibly on the state of the cell itself). The result of this 'computation' is then stored back in the cell on the next 'clock' period. In this way, all the cells are updated simultaneously in every cell, and the process repeats at each and every 'clock cycle'. In a 1D geometric CA, each cell has 2 neighbors, a 2D geometric CA has 8 neighbors, a 3D geometric CA has 26 neighbors, a 4D has 80 neighbors, and a 5D has 242, and so on. In general, if $C_D$ is the number of neighbors of an Nth dimensional geometric Cellular Automaton, and $N_{D-1}$ is the number of neighbors of the next lower N-1th dimensional geometric Cellular Automaton, then: $C_D = C_{D-1} + C_{D-1} + C_{D-}$ $_1$ + 2, or $C_D = 3 C_{D-1} + 2$, which is the number of neighbors of an Nth dimensional geometric CA expressed in terms of the number of neighbors in the next lowest space.

In EMQG, the geometric CA model chosen that best fits our universe is a simple geometric 3D CA, where each cell has 26 neighboring cells. Your first impression might be that the correct geometric CA model would be a 4D geometric CA, so that it is directly compatible with relativistic space-time. There are several problems with this approach, however. First 3D space and time have to be united in this CA model, which is not easy to do. More difficult still, is that 4D space-time is considered to be relative, and in some cases it is even curved like on the earth or in accelerated frames. This means that on the earth for an observer in a falling reference frame, there exists a flat 4D space-time, while a nearby stationary observer also stationed on the earth observes a curved 4D space-time. Furthermore, the 4D space-time curvature is directional near the earth. Curvature varies along the radius vectors of the earth, but does *not* vary parallel to the earth's surface (over small distances). It is hard to see how one constructs a CA model that is capable of this type of behavior.

To resolve these problems we proposed, in EMQG, *two separate* and different space and time structures in our universe (ref. 1). First, there is the relativistic 4D space-time that is measured with our instruments composed of matter, and which is influenced by accelerated motion and by gravity. Secondly, there is an absolute 'low level' 3D cellular automaton space, and separate 'time' that is not directly accessible to us by measurement. This exists at the lowest scale of distance and time, and is strictly a cellular automaton process involving information patterns. 3D space comes in the form of cells, which are sites for the storage of numbers. This automatically quantizes space.

Time is represented by the unidirectional evolution of the numeric state of the CA through the local rules, and by a regular clock period or 'clock cycle'. The numeric state of the CA changes state on each and every 'clock' pulse (an external synchronizer of the CA), which can be thought of as a primitive form of CA time ('clock cycles' should not to be confused with our clocks, or our measure of time). Thus in EMQG theory our low-level space and separate time is simply a 3D geometric CA. This model restores utter simplicity to the structure of our universe. Everything is deterministic, and the future evolution of the universe can in principle be determined, if the exact numeric state of the cellular automaton is known at a given point in time, along with the common set of rules that governs the behavior of each and every cell. Of course this is not possible in actual practice.

At what distance and time scale is the 3D geometric CA computer structure revealed, with respect to our measurement of distance and time? Assuming that the quantization scale corresponds to the Plank Scale (ref. 10) the number of cells per cubic meter of space is astronomically large; roughly $10^{105}$ cells. (*In EMQG, the quantization scale is actually much finer than the Plank Scale of distance and time*). Remember that all the cells in the universe are all updated in one single 'clock' cycle! This is a massive computation indeed! The number of CA 'clock' pulses that occur in one of our seconds is a phenomenal $10^{43}$

'clock cycles' per second (based on the assumption that the Plank distance of 1.6 x 10$^{-35}$ meters is the rough quantization scale of space, and the Plank time is 5.4 x 10$^{-44}$ seconds). Because of the remotely small distance and time scales of quantization, we as observers are very far removed from the low-level workings of our CA computer. Why do humans exist at such a large scale, as to be remotely removed from the CA cell structure? The simple answer to this is that life is necessarily a complex process! Even an atom is remarkably complex. A lot of storage locations (cells) are required to support the structure of an atom, especially in light of the complex QED processes that are going on. It is not possible to assemble anything as complex as life forms, without using tremendous numbers of atoms and molecules, and therefore many cells.

Recent studies of chaos theory and complexity theory teaches us that simple rules can lead to enormous levels of complexity. We can even see this in a simple 2D geometric CA called the game of life. Being a 2D geometric CA, there are 8 neighbors for each cell, which forms a primitive geometric 2D space, which can be viewed on a computer screen. Here the rules are very simple. The rules for the game of life are summarized below:

- If a given cell is in the one state, on the next clock pulse it will stay one when it is surrounded by precisely two or three ones among it's eight neighbors on a square lattice. If the cell is surrounded by more than three neighbors, it will change to zero; if fewer than two neighbors have a one, it changes to zero.

- If a given cell is in the zero state, on the next clock pulse: it will change to a one when surrounded by precisely three ones, otherwise for any other combination of neighbor states the cell will remain a zero.

These rules are even simple enough for a child to understand, yet the game of life leads to an endless number of different patterns, and to a significant complexity in many of the processes found. For example, some of the things we see are gliders, puffers, guns, 'oscillating' particles with different rates of translation, and spontaneous particle emission from oscillating patterns. We have even seen a pattern that somewhat resembles a particle exchange process (although it is far from a perfect example).

Given that the CA is a 2D geometric CA, how many cellular automata can be constructed from all the possible rules for a 2D CA? This number is unimaginably large. For simple binary cells, with 8 neighboring cells there are 8+1 cells that influence a given cell (previous state of a cell can influence it's next state), which leads to $2^{512}$ possible binary combinations or approximately $10^{154}$ different CA's, of which the game of life is but one. In general, for an Nth Dimensional Geometric CA with (m) neighbors, there are $2^k$ possible rules available for the Cellular Automaton, where k = $2^{(m+1)}$. Assuming our universe is a simple 3D geometric CA, then there are $2^{134,217,728}$ possible rules to choose from! You can give up trying to find the rules that govern our universal CA by simple trial and error.

In the early 1900's, Max Plank proposed (ref. 10) a set of fundamental scales based on his newly discovered quantum of energy h (E = hv). Three fundamental constants G (Gravitational Constant), h (Plank's Quantum of Action), and c (the speed of light) were assembled in a set of equations that define natural physical units independent of any man made units. The Plank length is $1.6 \times 10^{-35}$ meters and the Plank time is $5.4 \times 10^{-44}$ seconds. The Plank length has often been suggested as the fundamental quantization scale of our universe. This suggests that the 'size' of a cell in the CA computer is the Plank length, and that the Plank time is the period of the cellular automaton 'clock' (the smallest possible time period for a change in the state of the CA). Note: The cells actually have no real size, since they represent storage locations for numbers. Instead, a cell represents the smallest possible distance that you can increment when you move from place to place.

There is currently evidence to suggest that the plank units of distance and time somehow represent the quantization scale of space-time itself. For example, when the strength of the four forces of nature are compared, they become comparable in strength near the plank scale. In CA theory, quantization is automatic and no extra steps are required! There are also some other physical units that can be derived from the three fundamental constants listed above which includes: Plank Energy, Plank Temperature, Plank Mass, Plank Speed, and Plank Wavelength, which also represent fundamental universal limits to these parameters.

All physical things result from CA processes, including space, time, forces and matter (particles). Although, the exact rules of the CA that is our universe is unknown at this time, some very general physical conclusions can be drawn from the CA model. For example, matter is constructed from elementary particles, which move in space during some period of time. Particles interact via forces (exchange particles) which bind particles together to form atoms and molecules.

Quantum field theory tells us that forces are also particles called vector bosons, which are readily exchanged between matter particles (fermions), and that these exchanges cause momentum changes and accelerations that we interpret as forces. Elementary particles and forces on the CA consist of oscillating information patterns, which are numbers changing state dynamically in the cells of the CA. These numerical information patterns roam around from cell to cell in given directions. The shifting rates of a particle or information pattern, relative to some other particle, is interpreted by us as the state of relative motion of the particle.

We have seen that particles can interact with other particles by exchanging particles, which are also information patterns. Exchange particles are readily emitted at a given fixed rate by the source particle, and absorbed by target particle. Similarly, the destination particle is also an emitter of exchange particles, and these are absorbed by the source particle When force particles are absorbed, the state of the internal information pattern changes, which results in a change in the particle momentum.

The result of this process is that the shifting rates or motion of the matter particle changes by undergoing a positive or negative acceleration with respect to the source. This is what we observe as a fundamental unit of force. Of course, when we observe forces on the macroscopic scale, the astronomical number of particle exchanges occurring per second blurs the 'digital' impact nature of the force exchange, and we perceive a smooth force reaction. All 'motion' is relative in CA theory, since all cells are identical and indistinguishable. In other words, we cannot know the specific cell locations that a particle occupies.

The closest element of the CA model that correspond to our space and time is the empty cells and the clock cycles that elapse. However, this correspondence is not exact. The cells, which are storage locations for numbers, form the lowest level of the concept of space. Since the information patterns can roam freely in various directions that are determined by the dimensionality of the CA, we interpret this freedom of motion as space. Similarly, while matter patterns are in motion, a definite time period elapses on the CA in the form of a specific number of CA 'clock cycles'.

We can only sense the elapse of time when matter is in motion, which is signified by the changing state of the CA. The ultimate cause of change or motion is the CA 'clock cycle', and the common rules that governs all cells. However, it is important to realize that the internal clock required for the CA to function is not the same as our measure of time in our universe. Our time is based on physical phenomena of matter interactions only. This, in fact, is the origin of much confusion of the nature of time in physics. Physicist generally consider time measured by clocks or other measuring instruments to be a basic element of physical reality.

We can now see that CA theory restores a great unity to physics. Where there used to be different phenomena described by different physical theories, now there is basically only one theory. Where space and time were considered to be fundamental aspects of our reality, we find that these yield to a more basic and fundamental concept. Furthermore, CA theory is not only able to describe the way the universe works, but has the potential to allows us to understand *how* it works in great detail.

Is there any real physical evidence to support the idea that our universe really is a Cellular Automaton computer simulation? The following sections will provide some rather speculative, and sometimes circumstantial, evidence to support this position. First we must introduce another important concept in understanding the quantum nature of Newton's laws: the virtual particles of the quantum vacuum. We will see that the quantum vacuum shares a common characteristic with most Cellular Automata simulations; namely the unavoidable swarm of processes filling all cells with what appears to be totally random activity.

## 4. THE QUANTUM VACUUM AND IT'S RELATIONSHIP TO CA THEORY

*Philosophers:    "Nature abhors a vacuum."*

By definition of the word 'vacuum', the vacuum is supposed to be totally empty. One makes a vacuum by removing all the gases and matter from a container. After this is done, you might conclude the vacuum is empty. In fact, the vacuum is far from empty! As we said, in order to make a complete vacuum one must remove all matter from an enclosure. However, this is still not good enough. One must not forget to lower the temperature down to absolute zero. This is required in order to remove all thermal electromagnetic radiation, which come in the form of photon particles that would spoil the perfect vacuum.

However, Nernst correctly deduced in 1916 (ref. 32) that empty space is still not completely devoid of all radiation after this is done. He predicted that the vacuum is still permanently filled with an electromagnetic field propagating at the speed of light, called the zero-point fluctuations (or sometimes called vacuum fluctuations). This was later confirmed by the full quantum field theory developed in the 1920's and 30's. Later, with the development of QED, it was realized that all quantum fields should contribute to the vacuum state, like virtual electrons and positron particles, for example. Of course Newton was totally unaware of the nature of the quantum vacuum in his day.

Modern quantum field theory actually requires that the perfect vacuum to be teeming with activity, as all types of quantum virtual fermion (matter) particles and virtual bosons (force particles) from the various quantum fields appear and disappear spontaneously. These particles are called 'virtual' particles because they result from quantum processes that have short lifetimes, and are generally undetectable. One way to look at the existence of the quantum vacuum is to consider that quantum theory forbids the absence of motion, as well as the absence of propagating fields (exchange particles).

According to QED, the quantum vacuum consists of the virtual particle pair creation/annihilation processes (for example, electron-positron pairs), and the zero-point-fluctuation (ZPF) of the electromagnetic field (virtual photons). The existence of virtual particles of the quantum vacuum is essential to understanding the famous Casimir effect (ref. 11), an effect predicted theoretically by the Dutch scientist Hendrik Casimir in 1948. The Casimir effect refers to the tiny attractive force that occurs between two neutral metal plates suspended in a vacuum. He predicted theoretically that the force 'F' per unit area 'A' for plate separation D is given by:

$$F/A = - \pi^2 h c / (240 D^4) \quad \text{Newton's per square meter} \quad \text{(Casimir Force 'F')} \quad (4.1)$$

This minute force can be traced to the disruption of the normal quantum vacuum virtual photon distribution between two nearby metallic plates. Certain photon wavelengths (and therefore energies) in the low wavelength range are not allowed between the plates, because these waves do not 'fit'. This creates a negative pressure due to the unequal energy distribution of virtual photons inside the plates as compared to outside the plate region. The pressure imbalance can be visualized as causing the two plates to be drawn together by radiation pressure. Note that even in the vacuum state, virtual photons carry energy and momentum.

Recently, Lamoreaux made (ref. 12) accurate measurements for the first time on the theoretical Casimir force existing between two gold-coated quartz surfaces that were spaced 0.75 micrometers apart. Lamoreaux found a force value of about 1 billionth of a Newton, agreeing with the Casimir theory to within an accuracy of about 5%.

The virtual particles of the quantum vacuum is central to our understanding of the quantum processes behind Newton's laws. We therefore present a brief review of some of the theoretical and experimental evidence for the existence of the virtual particles of the quantum vacuum:

(1) The extreme precision in the theoretical calculations of the hyper-fine structure of the energy levels of the hydrogen atom, and the anomalous magnetic moment of the electron and muon; these calculations are based on the existence of virtual particles within the framework of QED theory. These effects have been calculated to a very high precision (approx. 10 decimal places), and these values have also been verified experimentally to an unprecedented accuracy. This is a great achievement for QED, which is essentially a perturbation theory of the electromagnetic quantum vacuum. Indeed, this is one of greatest achievements of theoretical physics.

(2) Recently, vacuum polarization (the polarization of electron-positron pairs near a real electron particle) has been observed experimentally by a team of physicists led by David Koltick (ref. 33). Vacuum polarization causes a cloud of virtual particles to form around the electron in such a way as to produce an electron charge screening effect. This is because virtual positrons tend to migrate towards the real electron, and the virtual electrons tend to migrate away. A team of physicists fired high-energy particles at electrons, and found that the charge screening effect of this cloud of virtual particles was reduced the closer a particle penetrated towards the electron. They reported that the effect of the higher charge for the penetration of the electron cloud with energetic 58 giga-electron volt particles was equivalent to a fine structure constant of 1/129.6. This agreed well with their theoretical prediction of 1/128.5 of QED. This can be taken as verification of the vacuum polarization effect predicted by QED, and further evidence for the existence of the quantum vacuum.

(3) The quantum vacuum explains why cooling alone will never freeze liquid helium, no matter how low the temperature. Unless pressure is applied, the quantum vacuum energy fluctuations prevents it's atoms from getting close enough to trigger solidification.

(4) For fluorescent strip lamps, the random energy fluctuations of the virtual particles of the quantum vacuum cause the atoms of mercury (which are in their exited state) to spontaneously emit photons by eventually knocking them out of their unstable energy orbital. In this way, spontaneous emission in an atom can be viewed as being directly caused by the random state of the surrounding quantum vacuum.

(5) In electronics, there is a limit as to how much a radio signal can be amplified. Random noise signals are always added to the original signal. This is due to the presence of the virtual particles of the quantum vacuum as the real radio photons from the transmitter propagate in space. The vacuum fluctuations add a random noise pattern to the signal by slightly modifying the energy content of the propagating radio photons.

(6) Recent theoretical and experimental work done in the field of Cavity Quantum Electrodynamics (the observation of exited atoms surrounded by a conducting cavity) suggests that the orbital electron transition time for excited atoms can be affected by the state of the virtual particles of the quantum vacuum immediately surrounding the excited atom in a cavity, where the size of the cavity modifies the spectrum of the virtual particles.

In the weight of all this evidence, we believe very few physicists doubt the existence of the virtual particles of the quantum vacuum. Yet to us it seems strange that the quantum vacuum should barely reveal it's presence. It is strange that we only know about it's existence through some rather obscure physical effects. After all, the observable particles of ordinary real matter constitutes an absolutely minute fraction of the total population of virtual particles of the quantum vacuum at any given instant of time. In other words if we could observe the detailed numeric cellular automaton operation of our universe, the interactions of the virtual particles of the quantum vacuum are, by far, the most common process we would find. In fact, it would be extremely difficult to follow a real matter process such as a hydrogen atom, since the electron following it's orbital would be constantly destroyed and recreated by countless interactions with the quantum vacuum.

Instead, we believe that the quantum vacuum plays a ***much more prominent role*** in all of physics then is currently believed. We maintain that the effects of the quantum vacuum are felt in *virtually all* physical activities. In fact, Newton's three laws of motion can be understood to originate directly from the effects due to the virtual particles of the quantum vacuum.

Is there any relationships that exist between Cellular Automata model and the quantum vacuum? Recall that the quantum vacuum implies that almost ***all*** of empty space is filled with virtual particle processes. Through studies of simple 2D geometric CA's (such as the Conway's Game of Life), most random initial states or 'seed' patterns on the cells (and often from small localized initial patterns, with all the remaining cells being in the zero state) are observed to evolve into a complex soup of activity, everywhere. This CA activity is very much reminiscent of the activities believed to be happening in the quantum vacuum. In the game of life you can even see events that even look suspiciously like random 'particle' collisions, particle annihilation, and particle creation after a sufficiently long simulations. Of course this is not hard evidence that our universe is a vast CA, but it is suggestive. We now present some other (somewhat) circumstantial evidence to support the view that our universe is a vast CA simulation. For those only interested in Newton's laws may skip the next section, and go directly to section 6 on Quantum Inertia.

## 5. EVIDENCE FOR BELIEVING THAT WE LIVE IN A VAST CA

Is there any evidence, direct or indirect, that we are indeed living in a universe that is a vast Cellular Automaton simulation? Here we list some of the important circumstantial evidence for believing that this is true:

### (A) THE BIG BANG (START OF SIMULATION AT T=0) AND CA THEORY

*"I want to know how God created this world (Universe)"*                    *- A. Einstein*

If the universe is a vast CA computer simulation, then it stands to reason there must have been a point where the simulation was first started. This occurred ≈15 billion years ago (our time), according to the standard hot big bang theory. It is important to realize that the creation of the numeric state of our universe, if it were to be done *now* in a single step or single act of creation, would be *very* difficult to accomplish. All the galaxies, stars, and planets, and the wide variety of life forms must be specified for all the cell states in the cellular automaton, which for our universe is something on the order of $10^{100}$ cells per cubic meter of space!

We currently believe that our universe contains on the order of a few billion galaxies, and many of these galaxies have something on the order of 100 billion stars in it. Currently, there is also evidence for the possibility that a certain percentage of these stars have one or more planets circling around them. Each star and planet has it's own unique orbit, chemical composition, temperature, rotation rate, size, atmosphere, landscape and possibly even life forms. In the process of creating our universe, it is far more economical to start with just the "right" rules of the cellular automaton so that stars and planets are the natural byproducts of the evolution of the CA over many, many clock cycles (and possibly the evolution of life itself).

Basically you start with the right CA structure, the right rules, and the correct initial cell patterns and let the natural evolution of the CA run its due course. It is also more "interesting" to start this process, and than "see" what comes out of it after a lot of computation. In fact, that is what the purpose of computer is anyway. The purpose of our CA computer universe is to compute our universe! CA theory absolutely requires that our universe be an evolutionary process, with a simple beginning. Of course many cells and clock cycles would be required, and that is just what we have now! With $10^{100}$ cells per cubic meter of space, and $10^{43}$ 'clock cycles' per second of our measure of time and 10 billion years of operation, almost anything is possible!

## (B)  WHY OUR UNIVERSE IS MATHEMATICAL IN NATURE

*"Why is it possible that mathematics, a product of human thought that is independent of experience, fits so excellently the objects of physical reality."*

*- Albert Einstein*

It is clear to physicists that all the known laws of physics are mathematical in nature. Many physicists like Einstein, for example, have commented on this mysterious fact as the quote above signifies. Yet, no good explanation has been given as to why this should be so for our universe. This fact is made even more mysterious when one considers that mathematics is strictly an invention or byproduct of intellectual activity.

In a sense, mathematics is like art and music. For example, the mathematical concepts of infinity, the imaginary numbers, and the Mandelbrot set in the complex plane are all mathematical objects that are invented by mathematicians. In mathematics, you start with virtually any set of self-consistent axioms, and formulate new mathematics as you please. Mathematics is strictly a *creative* process. Yet, our universe definitely operates in a mathematical way. Every successful physical theory has been formulated in the language of pure mathematics, and a good theory can even predict new phenomena that was not expected from the original premises of the theory.

If the universe is a cellular automaton, then there is a clear explanation as to why the universe is mathematical in nature. Quite simply put, everything in our universe is numerical information, which is governed by mathematical rules that specify how the numbers change as the computation progresses. In short, "***the universe is numbers***"**,** as was once proclaimed by the great Greek philosopher and mathematician Pythagoras. The design of the cellular automaton must have required intelligence, which was applied to the cellular automaton in the form of the mathematical rules for the cells. CA theory claims that all the laws of physics that we know today are mathematical descriptions of the underling, discrete mathematical nature of the numeric patterns that are present in our universal cellular automaton. Fredkin (ref. 34) once proposed that the universe should be modeled with a single set of cellular automaton rules, which will model all of microscopic physics exactly. He called this CA 'Digital Mechanics'. The laws of physics in this form are discrete and deterministic, and would replace the existing differential equations (based on the space-time continuum) for modeling all phenomena in physics.

The push to discover the theory of everything (or simply *The Theory* as it is now known) should not be looking for a set of partial differential equations which simply incorporates relativity and quantum field theory. Instead, we should be looking for the correct structure of the universal CA, and the corresponding set of logical rules that govern it's operation.

### (C) QUANTUM PARTICLES IN SAME STATE ARE INDISTINGUISHABLE

*" Common sense is the layer of prejudice laid down in the mind prior to the age of 18"*
*A. Einstein*

A particle physicist once remarked that elementary particles behave more like mathematical entities rather than familiar point-like particles. Particles are able to transform from one species type to another. Particles seem to be spread-out in some sort of oscillatory wave, and at other times they seem like point-like objects. Particles can be readily annihilated and created. Electrically charged electrons emit a never ending stream of photons, without degradation to the original electron. None of these processes seem familiar from our everyday experience of how matter ought to behave. One of the most unfamiliar of all quantum particle attributes is principle of indistinguishability of particles in the same quantum state.

According to quantum mechanics, electrons that are in the same quantum state (or having the same quantum numbers) are absolutely identical, and indistinguishable from each other. You cannot mark one electron so that it is different than another. An electron is currently described by quantum mechanics as a particle with quantum numbers like: mass, charge, spin, position, and momentum, which are represented as numbers in the wave function of the electron. It is these quantum properties alone tell you all there is to know about the electron. In other words, the electron has no size or shape. Quantum mechanics has ruled many classical or 'mechanical' type models to help us 'visualize' what an elementary particle really is.

Equality is strictly a mathematical concept. In mathematics, the equality of '1+1=2' is exact. However in classical physics no two marbles can be constructed to be exactly the same. When it comes to elementary particles however, two quantum particles can be *exactly* the same. Two electrons in the same state of motion (and same spin) are absolutely identical and indistinguishable. The cellular automaton model explains this remarkable fact simply by stating that the two electrons in the same state have *exactly* the same numeric information pattern, and are thus described by the same quantum wave function. Therefore they truly are mathematically identical. In fact, when constructing a universe it is very desirable to have building blocks that are identical, and exactly repeatable, so that large and predictable structures can be easily formed.

### (D) WHY IS THE LIGHT SPEED THE MAXIMUM SPEED YOU CAN GO?

Special Relativity theory is founded on Newton's laws and these two basic postulates:

*(1) The velocity of light in a vacuum is constant and is equal for all observers in inertial frames (inertial frame is one in which Newton's law of inertia is obeyed).*

*(2) The laws of physics are equally valid in all inertial reference frames.*

The special theory of relativity implies that the speed of light is the limiting speed for any from of motion in our universe. Furthermore, light speed appears constant no matter what inertial frame an observer chooses. However, nowhere in special relativity theory or any other theory that we are aware of, is there an explanation as to why this might be so. It is simply a postulate, based on physical observations such as the Michelson-Morley experiment. The second postulate also implies that there are no experiments that can be performed that will reveal which observer is in a state of 'absolute rest'. The second law alone is an alternative way of stating the Galilean transformations, which are an extension of Newton's laws of motion. With the addition of postulate 1, we are led to the Lorentz Transformations, which replace the Galilean transformations. The Lorentz transformation are the reason Newton's laws of motion break down at extreme velocities

The second postulate of special relativity states that the laws of physics are equally valid in all inertial reference frames. Stated in a weaker form, there are no preferred reference frames to judge absolute constant velocity motion (called inertial frames). This latter form is easily explained in CA theory, by remembering that all cells and their corresponding rules in the cellular automata are absolutely identical everywhere. Motion itself is an illusion, and really represents numeric information transfers from cell to cell. To assign meaning to motion in a CA, one must relate information pattern flows from one numeric pattern group with respect to another group, since the actual cell locations are totally inaccessible by experiment. Therefore motion requires reference frames. Unless you have access to the absolute location of the cells, all motion must remain relative in CA theory. In other words, there is *no* reference frame that is accessible by *experiment* that can be considered as the absolute reference frame for constant velocity motion.

In a cellular automaton, the clock rate specifies the time interval in which all the cells are updated, and acts as the synchronizing agent for the cells. Matter is known to consist of atoms and molecules, which themselves consist of elementary particles bound together by forces. An elementary particle in motion is represented in CA theory by a shifting numeric information pattern, that is free to 'roam' from cell to cell. Recall that space consists of cells or storage locations for numbers in the cellular automata, and particles (number patterns) freely 'move' in this cell space. From these simple ideas, it can be seen that there must be a maximum rate that numeric information patterns are able to achieve.

This is due to the following two reasons:

- First, there is fixed, constant rate in which cells can change state.
- Secondly, information can only be transferred sequentially, from one cell to adjacent cell, and only by one cell at a time per clock cycle.

These are simply due to the limitations of the structure of cellular automata computer model. Recall that the CA provides the most massively parallel computer model known. It is the CA's high degree of parallelism that is responsible for these limitations, because a particular cell state can only be affected by its immediate neighbors. Information can only evolve after each 'clock' period, and numeric information patterns can only arrive at a distant location by shifting from cell to adjacent cell. This result in a *definite* **maximum speed limit** for transfer of information patterns (particles) on the CA This maximum speed limit might represent light velocity, which is the fastest speed any particle can move.

*NOTE: According to EMQG this speed limit is actually the 'low-level' light velocity, defined as the velocity of photons in between scattering encounters with the electrically charged virtual particles, measured in the absolute CA units of measure (cells and clock cycles). The electrical scattering process with the charged virtual particles of the quantum vacuum reduces the 'low-level' light velocity to the value we observe for light in the vacuum. This process is somewhat like how transparent water reduces the velocity of light (in the vacuum), which is also responsible for the index of refraction of water. Reference 1 describes the details of this process.*

This maximum speed limit can be calculated if the precise quantization scale of space and time on the cellular automata level is known. Let us assume for now that the quantization of space and time corresponds *exactly* to the plank distance and the time scales (not true in EMQG theory). This means that the shifting of one cell represents a change of one fundamental plank distance $L_P$: $1.6 \times 10^{-35}$ meters, and that the time required for the shift of one cell is one fundamental plank time $T_P$: of $5.4 \times 10^{-44}$ seconds. Let us further assume that a photon represents the fastest of all the information patterns that shifts around in the CA. In fact, we propose that the photon information pattern is *only* capable of shifting one cell per clock period, and not at any other rate, and therefore exits at one speed with respect to the cells. The value for the speed of light can then be derived simply as the ratio of (our) distance over (our) time for the information pattern transfer rate. The maximum information transfer velocity is thus:

$$V_P = L_P / T_P = 3 \times 10^8 \text{ meter/sec} = c \qquad (9.1)$$

Therefore $V_P$ is equal to 'c', the speed of light. The velocity of light can also be expressed as one plank velocity, which is defined in units of plank length divided by plank time. (There are plank units for mass, temperature, energy, etc. as detailed in ref. 10).

Thus the fastest rate that the photon can move (shift) is an increment of one cell's distance, for every 'clock cycle'. If two or more 'clock cycles' are required to shift information over one cell, then the velocity of the particle is lower than the speed of light.

To summarize, in cellular automata theory the maximum speed simply represents the *fastest* speed in which the cellular automata can transfer information patterns from place to place. Matter (particles) are information patterns in the cellular automata, which occupies a number of cells. The cells themselves provide a means where information can be stored or transferred, and this concept corresponds to what we call the 'low level' CA space. 'Low level' CA time corresponds to the time evolution of the state of the cellular automata, which is governed by the 'clock period'.

To put it another way, the rate of transfer of information in any cellular automata is limited, and infinite speeds are simply not possible. Of course, this rules out action at a distance, which is why we consider CA theory to be manifestly compatible with special relativity.

It is interesting to note that in the famous 2D Geometric CA, called Conway's game of life, there exists a stable, coherent 'L' shaped pattern commonly known to computer scientists as a 'glider' pattern. This pattern is always contained in a 3 x 3 cell array, and the glider completes a kind of an internal 'oscillation' in four clock cycles. Thus in four clock cycles it returns to it's initial 'L' shaped starting pattern. This glider travels in 2D cell space, at only *one fixed speed*! It does not slow done or speed up in any circumstance. It is also the fastest moving pattern that we are aware of in the game of life. The glider particle in some sense resembles the photon particle in our universe. It has an internal oscillation, and it only moves at one fixed velocity. However, the similarity ends here because in the game of life, the glider only moves in *four* fixed directions.

We now move on to describe quantum inertia theory, which is part of EMQG theory and is of central importance to understanding Newton's three laws of motion. Quantum Inertia directly relates to gravitational interactions as well, and as such provides a basis for understanding why Newtonian inertial mass is equivalent to Newtonian gravitational mass. The reason for this equivalence has been a great mystery in physics since Newton's time. It has even escaped explanation by Einstein when he used mass equivalence as the basis for his famous (strong) equivalence principle. Here **all** physical processes that occur (in a small volume of space) on the earth are equivalent to the same physical process in a rocket accelerating at 1g (this would include, for example, the way water boils). One of the strongest points of EMQG is that mass equivalence can be seen to result from a common quantum vacuum process, occurring in both accelerated motion and under a gravitational field.

6. **THE QUANTUM THEORY OF INERTIA**

**"Under the hypothesis that ordinary matter is ultimately made of subelementary constitutive primary charged entities or 'partons' bound in the manner of traditional elementary Plank oscillators, it is shown that a heretofore uninvestigated Lorentz force (specifically, the magnetic component of the Lorentz force) arises in any accelerated reference frame from the interaction of the partons with the vacuum electromagnetic zero-point-field (ZPF). ... The Lorentz force, though originating at the subelementary parton level, appears to produce an opposition to the acceleration of material objects at a macroscopic level having the correct characteristics to account for the property of inertia."**
                                                                                                       **- B. Haisch, A. Rueda, H. E. Puthoff**

We have seen that CA theory absolutely demands that all global phenomena must be explainable from the strictly local rules (or program) governing the CA. Therefore physical phenomena such as acceleration and gravity should somehow originate from small-scale particle interactions dictated by the rules of the CA. Recall that CA theory is based on the local rules that apply to the local cellular neighborhood, and these rules are repeated on a vast scale for all the cells in the CA.

However, many of our existing physical theories are global principles of nature. This includes Newtonian inertia and Newtonian universal gravitation. In EQMG both Newtonian inertia and gravity have a detailed, particle level explanation based on the local "conditions" at the neighborhood of a given mass particle, and therefore compatible with the philosophy of a cellular automata theory, and with the principle of strict locality in special relativity.

In 1994 a new theory of inertia (ref. 5) was proposed by Haisch, Rueda, and Puthoff which is known here as the HRP Inertia. This event marked a revolution in the way we think about inertia. They argued that inertial force originates from the activity of the virtual particles that fills the perfect vacuum surrounding a mass. They suggested that it is this ever-present sea of energy that resists the acceleration of a mass, and so creates the Newtonian force of inertia. They have found a model of inertia that is traceable to the particle level, which is manifestly compatible with CA theory. Inertia is now the result of quantum particle interactions.

Haisch, Rueda, and Puthoff have come up with a new version of Newton's second law: F=MA. As in Newton's theory, their expression has 'F' for force on the left-hand side and 'A' for acceleration on the right. But in the place of 'M', there is a complex mathematical expression tying inertia to the properties of the vacuum (ref. 5). They found that the fluctuations in the vacuum (virtual photons) interacting with the charge particles of matter in an accelerating mass give rise to a magnetic field, and this in turn, creates an opposing force to the motion. Thus electrical forces or photon exchanges are ultimately responsible for the force of inertia!

They reasoned that the more massive an object, the more 'partons' it contains; and the more partons a mass contains means more individual (small) electrical forces from the vacuum are present and the stronger the reluctance to undergo acceleration. But, when a mass is moving at a *constant* velocity, inertia disappears, and there is no resistance to motion in any direction as required in special relativity.

In HRP theory inertia is caused by the magnetic component of the Lorentz force which arises between what the author's call 'parton' particles in an accelerated reference frame interacting with the background vacuum electromagnetic zero-point-field (ZPF). The author's use the old fashion term originated by Feynman called the 'parton', which referred to the elementary constituents of the nuclear particles such as protons and neutrons. It is now known that Feynman's partons consists of electrically charged quarks (which also posses color charge), and that the proton and neutron each contain three quarks of two types: called the 'up' and 'down' quarks. Although the neutron is electrically neutral, the constituent quarks are electrically charged, and the sum of the charges is zero.

We have found it necessary to make a *small* modification of HRP Inertia theory in our investigation of the principle of equivalence. In EMQG, the *modified* version of HRP inertia is called "Quantum Inertia", or QI. In EMQG, a new elementary particle is required to fully understand inertia, gravitation, and the principle of equivalence. **All** matter, including electrons and quarks, must be made of nature's most fundamental mass unit or particle which we call the 'masseon' particle.

The masseon particle posses one fixed, smallest possible quanta of gravitational 'mass charge'. The masseons also carry one fixed, smallest quanta of electrical charge as well, of which it can be either positive or negative. Masseons exist in particle or anti-particle form (called the anti-masseon). The masseon can also appear at random in the vacuum as virtual masseon/anti-masseon particle pairs of opposite electric charge. The earth consists of ordinary masseons (no anti-masseons), of which there are equal numbers of positive and negative electric charge varieties. The masseon particle model will be elaborated later. Instead of the 'parton' particles (that make up an inertial mass in an accelerated reference frame) interacting with the background vacuum electromagnetic zero-point-field (ZPF), we postulate that the real masseons (that make up an accelerated mass) interacts with the surrounding, virtual masseons of the quantum vacuum, electrically. However, the detailed nature of this interaction is still not known at this time. For example, why is it that for constant velocity motion the forces add to zero, but when acceleration is introduced the forces add up to Newton's inertial force? Since we still do not know the answers to these questions, we treat the Quantum theory of Inertia as a postulate of EMQG theory.

Quantum Inertia is deeply connected with our version of quantum gravity called ElectroMagnetic Quantum Gravity (EMQG). EMQG explains why the inertial mass and gravitational mass are identical in accordance with Einstein's weak equivalence principle. The weak equivalence principle translates to the simple fact that the mass (m) that measures the ability of an object to produce (or react to) a gravitational field ($F=GMm/r^2$) is the same as the inertial mass value that appears in Newton's F=MA.

In EMQG, this is not a chance coincidence. Instead, equivalence follows from a deeper quantum process occurring inside a gravitational mass involving quantum interactions with the quantum vacuum that are *very similar* in nature to the same quantum interactions involved in an accelerated mass.

In this view, the virtual particles in the quantum vacuum can be considered to form a kind of absolute reference frame for accelerated motion only. This absolute frame is simply represented as the resultant average acceleration vector of a mass with respect to the average acceleration vector of the quantum vacuum as a whole. This quantum vacuum reference frame can be used to gauge absolute acceleration. We do not need to measure our motion with respect to this frame in order to confirm that a mass is accelerated. We simply need to measure if an inertial force is present from the vacuum. Because mass turns out to be a complex phenomena, we present our definitions of inertial mass, gravitational mass, and a new form of mass called the gravitational 'mass charge' in the next section.

## 7. THE BASIC MASS DEFINITIONS OF EMQG

*"Subtle is the lord…"*

*- Einstein*

Based on quantum inertia and the quantum principle of equivalence (ref. 1), EMQG proposes three ***different*** mass definitions for an object possessing mass, which are listed below:

- **INERTIAL MASS** is the measurable mass defined in Newton's force law F=MA. This is considered as the absolute mass in EMQG, because it results from force produced by the relative (statistical average) acceleration of the charged virtual particles (the virtual masseons) of the quantum vacuum with respect to the electrically charged particles that make up the inertial mass (the masseons). To some extent, the virtual particles of the quantum vacuum is equivalent to Newton's absolute reference frame. In special relativity, inertial mass is equivalent to the rest mass of a particle.

- **GRAVITATIONAL MASS** is the measurable mass involved in the gravitational force as defined in Newton's law $F=GM_1M_2/R^2$. This is what is measured on a weighing scale. This is also considered as the absolute mass of an object, and is (almost) exactly the same as inertial mass. The same quantum process responsible for mass in inertia is also (partly) responsible for gravitational mass.

- **LOW LEVEL GRAVITATIONAL 'MASS CHARGE'** which is the source of the pure gravitational force, and is defined as the force that results through the exchange of graviton particles between two (or more) quantum particles (masseon particles). This type of mass is analogous to 'electrical charge', where photon particles are exchanged between electrically charged particles (electrons). Note that this is very hard to measure directly, because it is masked by the background quantum vacuum electrical force interactions which dominates over the pure graviton force exchange processes in gravitational interactions on the earth.

It is also important to note that these three forms of mass are ***not, in general equal***! Gravitational 'mass charge' is our most fundamental definition of mass, and is basically a quantum process similar to electrical charge. However, the gravitational force that arises from the graviton exchanges directly, is very difficult to measure. It is almost impossible to untangle this low-level force from electrical force processes also occurring from the electrically charged virtual particles of the quantum vacuum.

Both inertial mass and gravitational mass involve this electrical force process with the virtual particles of the quantum vacuum. It turns out that the inertial mass is almost exactly the same as gravitational mass, but *not perfectly* equal due to a minor difference between quantum processes in accelerated and gravitational frames. Direct graviton exchanges between a test mass and earth upset equivalence, which are not present in accelerated frames.

All quantum mass particles (fermion particles, which are composed of masseons) have all three mass types defined above. But bosons (only photons and gravitons are considered in this work) have only the first two mass types. This means that photons and gravitons transfer momentum and *do* react to the presence of accelerated frames and to gravitational fields, but they do not emit or absorb gravitons (ref. 1). Gravitational fields also affect photons, and this effect is linked to the concept of space-time curvature (ref. 1, for full details). It is important to realize that gravitational fields deflect photons (and gravitons), but not by force particle exchanges directly (as is believed in conventional physics). Instead, it is due to the electrical scattering of photons with the electrically charged virtual particles of the quantum vacuum, where the quantum vacuum can be thought of as a moving, transparent fluid flow (like water scattering light in the famous Fizeau experiment, ref. 1).

Gravitational mass also involves the same 'inertial' electrical force component that exists for in an accelerated mass, thus revealing a *deep connection* between inertia and gravity. For a stationary mass on the surface of the earth, there exists a similar quantum vacuum process that occurs in Newtonian inertia, where now the roles of the real charged mass particles of the mass and the electrically charged virtual mass particles of the quantum vacuum are reversed. Now it is the electrically charged virtual particles of the quantum vacuum that are accelerating (downward), and the mass particles of the mass that are at relative rest.

The reason for the downward acceleration of the virtual mass particles of the quantum vacuum is the absorption of gravitons from the earth (all mass particles possess 'mass charge'). Thus inside gravitational fields there is a *hidden* accelerated motion. The hidden acceleration is the same as the readily visible acceleration in an accelerated mass. However, there are also direct graviton exchanges between the earth and the mass particles inside the stationary mass on the earth, and these graviton exchanges result in a slight imbalance in mass equivalence to be discussed later.

What is unique about EMQG theory as a quantum field theory of gravity is that gravity simultaneously involves *two boson exchange particles*! In gravity, *both* the electrical force (photon exchange process) and the pure gravitational force (graviton exchange process) interactions are occurring simultaneously for all gravitational interactions. Graviton particles originate from all the mass particles (masseons) in the earth in vast numbers, and are absorbed by any particle possessing mass. Since the quantum vacuum consists of virtual particles which possess mass (virtual masseons) the virtual masseons are in a state of free fall during their very short lifetimes. We have seen that the quantum vacuum originates an electrical force interaction with electrically charged particles that make up a mass when there is a state of relative acceleration. Therefore the falling vacuum causes a stationary test mass on the earth to exert a force against the earth's surface, which we call the weight of that mass. Furthermore, we can now easily see why the weight of a mass on the earth is equal to the inertial mass of the same object on the floor of a rocket accelerating at 1g.

As we hinted for a stationary gravitational test mass on the earth, there is a direct graviton exchange process (which is not present in accelerated reference frames) occurring directly between the earth and the test mass, which upsets the perfect equivalence of inertial and gravitational mass, where the gravitational mass is slightly larger than the inertial mass due to the extra force from the direct graviton exchanges. One of the consequences of this imbalance is that if a very large and a tiny mass are dropped simultaneously on the earth, the larger mass would arrive slightly sooner. This effect is a testable consequence of EMQG theory, and it might be possible to measure this in a laboratory (see the note at the end of section 12 for one proposed experiment).

## 8. MACH'S PRINCIPLE AND QUANTUM INERTIA

*"... it does not matter if we think of the earth as turning round on its axis, or at rest while the fixed stars revolve around it ... The law of inertia must be so conceived that exactly the same thing results from the second supposition as from the first."*

*E. Mach*

In the late 18$^{th}$ century, Ernst Mach proposed that the inertial mass of a body does not have any meaning in the absence of the rest of the matter in the universe. In other words, acceleration requires some other reference frame in order to determine accelerated motion. It seemed to Mach that the only reference frame possible was that of the average motion of all the other masses in the universe. This implied to Mach that the acceleration of an object must somehow be dependent on the sum total of all the matter in the universe. To Mach, if all the matter in the universe were removed, the acceleration, and thus the force of inertia would completely disappear since no reference frame is available to determine the actual acceleration.

Imagine a spinning elastic sphere that bulges at the equator due to the centrifugal force of the spin. This thought experiment is similar to Newton's thought experiment where a spinning bucket is filled with water, and the rotation causes the water around to bulge out around the rim. The question that Mach asked was how does the sphere 'know' that it is spinning, and therefore must bulge. If all the matter in the universe was removed, how can we be sure that it is really rotating? Therefore how would the sphere 'know' that it must bulge or not?

Newton's answer would have been that the sphere somehow felt the action of Newtonian absolute space. Mach believed that the sphere somehow 'feels' the action of all the cosmic masses rotating around it. To Mach, centrifugal forces are somehow gravitational in the sense that it is the action of mass on mass. To Newton, the centrifugal force is due to the rotation of the sphere with respect to absolute space. To what extent that Einstein's general theory of relativity incorporates Mach's ideas is still a matter of debate (ref. 35). EMQG (through the quantum inertia principle) takes a similar view as Newton, where Newton's absolute space is replaced by the virtual particles of the vacuum. Mach was never unable to develop a full theory of inertia based on his idea of mass affecting mass.

Mach's ideas on inertia are summarized as follows:

- A particle's inertia is due to some (unknown) interaction of that particle with all the other masses in the universe.

- The local standards of non-acceleration are determined by some average value of the motion of all the masses in the universe.

- All that matters in mechanics is the relative motion of all the masses.

Quantum inertia theory fully resolves Mach's paradox by introducing a new universal reference frame for gauging acceleration: the net statistical average acceleration vector of the virtual particles of the quantum vacuum with respect to the accelerated mass. In other words, the cause of inertia is the interaction of each and every particle with the quantum vacuum. Inertial force actually *originates* in this way. It turns out that the distant stars do affect the local state of acceleration of our vacuum here through the long-range gravitation force. Thus our local inertial frame is slightly affected by all the masses in the distant universe. However, in our solar system the local gravitational bodies swamp out this effect. (This long-range gravitational force is transmitted to us by the graviton particles that originate in all the matter in the universe, which will distort our local net statistical average acceleration vector of the quantum virtual particles in our vacuum with respect to the average mass distribution). Thus it seems that Mach was correct in saying that acceleration here depends somehow on the distribution of the distant stars (masses) in the universe, but the effect he predicted is minute. We now go on to describe Newton's laws of motion as a quantum process.

## 9. NEWTON'S THREE BASIC LAWS OF MOTION

We are now in a position to understand the quantum nature of Newton's three classical laws of motion. According to the standard treatment from textbooks (ref. 19) Newton's three laws of laws of motion can stated as:

***1. An object at rest will remain at rest and an object in motion will continue in motion with a constant velocity unless it experiences a net external force.***

***2. The acceleration of an object is directly proportional to the resultant force acting on it and inversely proportional to its mass. Mathematically: $\Sigma F = ma$, where 'F' and 'a' are the vectors of each of the forces and accelerations.***

***3. If two bodies interact, the force exerted on body 1 by body 2 is equal to and opposite the force exerted on body 2 by body 1. Mathematically: $F_{12} = -F_{21}$.***

Newton's first law explains what happens to a mass when the resultant of all external forces on it is zero. Newton's second law explains what happens to a mass when there is a nonzero resultant force acting on it. Newton's third law tells us that mechanical forces always comes in pairs. In other words, a single isolated mechanical force cannot exist. The force that body 1 exerts on body 2 is called the action force, and the force of body 2 on body 1 is called the reaction force.

According to Quantum Inertia theory, Newton's first two laws are the direct consequence of the electrical force interaction of the charged elementary particles of the mass interacting with the charged virtual particles of the quantum vacuum. Newton's third law of motion is a direct consequence of the fact that all forces are the result of a two way boson particle exchange process. Let us look at his three laws carefully in this framework.

**(1) Newton's First Law of Motion:**

In EMQG, the first law is a trivial result, which follows directly from the quantum inertia. First a mass is at relative rest with respect to an inertial observer in deep space. This means that the mass has a *net acceleration* of zero with respect to the electrically charged virtual particles of the quantum vacuum If no external forces act on the mass, the electrically charged elementary particles that make up the mass maintain a *net acceleration* of zero with respect to the electrically charged virtual particles of the quantum vacuum through the constant electrical force exchange process with the vacuum. This means that no change in velocity is possible (an acceleration of zero) and the mass remains at rest.

Secondly a mass has some given constant velocity with respect to an inertial observer in deep space. If no external forces act on the mass, the electrically charged elementary particles that make up the mass also maintain a ***net acceleration*** of zero with respect to the electrically charged virtual particles of the quantum vacuum through the electrical force exchange process (even though the mass moves at constant velocity with respect to an inertial observer. Again, no change in velocity is possible (zero acceleration) and the mass remains at the same original constant velocity.

**(2) Newton's Second Law of Motion (F=MA)**:

The second law basically *is* the quantum theory of inertia that we discussed above. Basically the state of *relative* acceleration of the electrically charged virtual particles of the quantum vacuum with respect to the electrically charged particles of the mass is what is responsible for the inertial force, through the tiny electrical force contributed by each mass particle undergoing an acceleration 'A', with respect to the net statistical average of the virtual particles of the quantum vacuum. This results in the property of inertia possessed by all masses. The sum of all these tiny electrical forces contributed from each charged particle of the mass from the vacuum is the source of the total inertial resistance force opposing accelerated motion in Newton's F=MA.

**(3) Newton's Third Law of Motion**:

According to the boson force particle exchange paradigm, a forces between two particles results from the two-way nature of particle exchanges. This includes mechanical contact forces, which result from the electrical forces between the outer electrons of one object acting against the other. Therefore the force that body 1 exerts on body 2 (called the action force), is the result of the emission of photon force exchange particles from the electrically charged particles that make up body 1, which are absorbed by the electrically charged particles that make up body 2, resulting in a mechanical force acting on body 2.

Similarly, the force of body 2 on body 1 (called the reaction force) is the result of the absorption of photon force exchange particles that are now originating from the electrically charged particles that make up body 2, and being absorbed by the electrically charged particles that make up body 1, again resulting in a mechanical force acting on body 1. An important property of electrical charge is the ability of particles to readily emit *and* absorb photon force exchange particles. Therefore body 1 is both an emitter and also an absorber of the photons. Similarly, body 2 is also both an emitter and an absorber of photons. This is precisely the reason that there is both an action and reaction force associated with two bodies undergoing mechanical forces.

For example, the contact forces (the mechanical force that Newton was thinking of when he formulated this law) that results from a person pushing on a mass and the reaction force from the mass pushing on the person is really the exchange of photon particles from the charged electrons bound to the atoms of the person's hand and the charged electrons bound to the atoms of the mass on the quantum scale. Therefore on the quantum level there is really is no direct contact of electrons. The hand gets very close to the mass, but the electrons do not actually touch. The electrons in one object are constantly exchanging photons with the electrons in the other object. The force exchange process works in both directions, and involves roughly equal numbers of photons. Therefore the two mechanical forces that result are equal and opposite in direction.

## 10. NEWTONIAN MOMENTUM AND THE INERTIAL FORCE

Quantum Inertia (QI) also provides a new understanding of Newtonian linear momentum. We argue that it is only ***inertial force*** that is truly a ***fundamental*** concept of nature, not momentum or conservation of momentum (and all forces in general). Momentum, which is defined by 'mv', is merely a bookkeeping value used to keep track of the inertial mass 'm' (defined by F/A) in the state of constant velocity motion 'v' ***with respect to another mass*** that it might collide with at some future time. In this way, momentum is a relative quantity.

Momentum simply represents bookkeeping information with respect to some other mass that it might encounter in a later, possible force interaction. This fits in with the fact that *inertial* mass cannot really be measured for constant velocity masses (in outer space for example, away from all other masses) without introducing some *test* acceleration of the mass. If a mass is moving at a constant velocity, there are *no* cumulative forces present from the vacuum. Furthermore, since momentum involves velocity, it requires some other inertial reference frame in order to gauge the velocity 'v'. The higher the velocity that a mass 'm' achieves, the greater will be the subsequent deceleration (and therefore the greater the subsequent inertial force interaction) during a later collision.

For example, if the velocity doubles with respect to a wall ahead, then the deceleration doubles in a later impact. Before doubling the velocity, the acceleration $a_0 = (v_0 - 0)/t$; and after doubling, $a = (2v_0 - 0)/t = 2a_0$. Therefore we find that $f = 2f_0$, the force required from the wall (assuming the time of collision is the same). Similarly, if the mass is doubled, the force required from the wall doubles, or $f=2f_0$. Recall that inertial force comes from the **opposition of the quantum vacuum to the acceleration of mass** (or deceleration as in this case). Similarly, the kinetic energy '$1/2mv^2$' of a mass moving at a constant relative velocity 'v', it is also a bookkeeping parameter defined as the product of <u>force</u> and the <u>time</u> that a force is applied. This quantity keeps track of the subsequent energy reactions that a mass will have when later accelerations (or decelerations) occur with respect to some other mass. It is important to remember that it is the electrical **quantum vacuum force** that acts against an inertial mass to oppose any change in its velocity that is truly the fundamental concept.

We therefore conclude that according to principles of QI theory, the inertial force is **absolute**. We also implied that acceleration *can* also be considered *absolute*. By this we mean that it is only when a mass 'm' undergoes an acceleration 'a' with respect to the net statistical average acceleration of the virtual particles of the quantum vacuum that an inertial force 'F' appears. Therefore one can infer the relative acceleration 'a' of a mass 'm' with respect to the quantum vacuum average acceleration simply by determining a=F/m. In other words, the presence of an inertial force reveals that there is an *absolute* acceleration with respect to the quantum vacuum.

## 11.   EMQG AND NEWTON'S UNIVERSAL LAW OF GRAVITY

We are now in the position to use the particle exchange paradigm in EMQG to understand Newtonian gravity. Gravity is one of the four basic forces of nature. The standard model of particle physics does not yet account for gravity as a pure exchange process successfully. The standard model only addresses the electromagnetic, weak and strong nuclear forces within the framework of quantum field theory. In quantum field theory, forces are thought to originate from the exchange of force particles (vector bosons) which are represented by the quanta of the associated classical field.

In EMQG *two* fundamental particle exchange processes working together are responsible for the force of gravity. The bosons involved are the familiar photon, and the graviton exchange particles. In EMQG, the photon and the graviton are almost identical in their physical properties, except for their relative strengths. The boson acts like the go between particle, shifting from cell to cell until it is absorbed by a destination particle. This transfers a force, without any action at a distance. We now examine the nature of the particle exchanges in more detail.

11.1   FORCES, PARTICLE EXCHANGES AND CELLULAR AUTOMATON

In order to understand gravity, let us briefly review QED and it's connections with CA. The theory that best describes photon particle exchanges and the quantization of the electromagnetic force field is Quantum Electrodynamics (QED). Here the charged particles (electrons, positrons) act upon each other through the exchange of force particles, which are called photons. According to classical electromagnetic theory, the force due to two charged particles decreases with the inverse square of their separation distance. The corresponding classical theory that describes this is Coulomb's electrical inverse square law: $F = kq_1q_2/r^2$, where k is a constant, $q_1$ and $q_2$ are the charges, and r is the distance of separation.

QED accounts for the Coulomb force law by postulating the exchange of photons between the electrically charged particles. The number of photons emitted and absorbed by a given charge (per unit of CA time) is fixed and is called the charge of the particle. Thus, if the charge doubles, the force doubles because twice as many photons are exchanged during the force interaction. This force interaction process causes the affected particles to accelerate either towards or away from each other depending on the charge value.

Certain 2D cellular automata exhibit behavior roughly resembling electrical charge. For example, in the famous 2D geometric CA known as Conway's game of life there exist a large variety of CA patterns types generally referred to as 'gun' patterns. Gun patterns are constantly emitting a steady stream of 'gliders' as they travel through 2D CA space. This emission process is constant and without any degradation whatsoever to the original 'gun' numeric information pattern. This process resembles the property of electron charge, where photons are constantly emitted without any degradation or change to the original electron.

QED explains why the strength of the electrical force varies as the inverse square of the distance of separation between the charges. Each charge sends and receives photons from every direction. But, the number of photons per unit area, emitted or received, decreases by the factor $1/4\pi r^2$ (the surface area of a sphere for a 3D geometric CA) at a distance 'r' due to the photon emission pattern spreading equally in all directions. Thus, if the distance doubles, the number of photons exchanged decreases by a factor of four.

This process can easily be visualized on a 3D geometric CA. Imagine that an electron particle is at the center of a sphere and sending out virtual photons in all directions. Imagine that another electron is on the surface of a sphere at a distance 'r' from the emitter, which absorbs some of these photons. The absorption of these photons causes an outward acceleration, and thus a repulsive force. If the charge is doubled on the central electron, there is twice as many photons appearing at the surface of the sphere, and twice the force acting on the other electron. This accounts for the linear product of charge terms in the numerator of the inverse square law. In QED, photons do not interact with each other (through a force particle exchange). As a result, in-going and out-going photons do *not* generally affect each other during the exchange process.

If you are not fully versed in modern quantum field theory you may question why two oppositely charged particles can be attracted to each other, while each is absorbing an exchange particle. On face value, classical thinking would imply that the momentum kick could only cause the particles to move apart! A typical QED textbook explains this fact by the mathematics of momentum transfer at the vertices (and the electron spin) of the associated Feynman fundamental process. Certainly, classical thinking cannot explain this process, nor can classical models explain why photons are constantly emitted without degradation to the original electron.

We believe the only way out of this quandary is to accept that an electron is an information pattern, which obeys the rules of the Cellular Automaton. We believe that the eight fundamental Feynman QED vertices are due to pure cellular automata processes, which are capable of transforming incoming particle information patterns into different outgoing particle information patterns. In other words, we believe that there must be a lot of *unseen CA activity* in the Feynman fundamental vertices, and that these details are hidden from the physicist because of the purely numeric aspect of this process

In 'Conway's game of life' CA, we discovered a CA pattern that is refered to as a 'loop' pattern, which evolves into something resembling a two particle exchange process. Two larger internal oscillating CA patterns are seen to move apart while 'glider' particles (the small, high-speed oscillating patterns reminiscent of photons) are exchanged. This pattern is something like a CA prototype pattern of a particle exchange process leading to a force.

However, we found that the 'loop' is not a perfect model, because the gliders are traveling in *only* four directions. Also, the particle exchange gives a constant velocity outward motion to source and destination particles, and *not an accelerated motion* as required. To our knowledge no one has found a perfect CA particle exchange process that looks identical to real physical particles exchanges in any Cellular Automata simulations. However, we believe that something like this is actually happening in vast numbers on the plank scale for real particle force interactions in our universe.

## 11.2 GRAVITATION ORIGINATES FROM GRAVITON EXCHANGES

For electrical forces, it is experimentally confirmed that the electrical force originating from two particles possessing electrical charge decreases with the inverse square of their separation distance, and is given by Coulomb's inverse square law: $F = Kq_{1}q_{2}/r^2$, where K is Coulomb's constant, $q_1$ and $q_2$ are the magnitude of the electrical charges, and r is the distance of separation. What if someone wanted to change this law slightly, so that the $r^2$ term is changed to something like $r^{1.999999145}$ or $r^{2.00000000786}$ for example? Would there be any good theoretical reason given against this, bearing in mind these new values would still give experimentally valid results. Since the $r^2$ term can only be arrived at by observation, it seems that a change like this cannot be ruled out, albeit a messy change. Also, the magnitude of the force depends on the product of the charges, and again this is only know experimentally to be true. No deeper theoretical reason is known why this should be exactly true. ***We maintain that according to the particle exchange paradigm, no changes to the $r^2$ term or the charge product term is possible in the inverse square law.***.

For gravitational force, it is experimentally observed that the force originating from two particles possessing mass also decreases with the inverse square of their separation distance, and is given by Newton's inverse square law: $F = Gm_1m_2 / r^2$, where G is Newton's universal gravitational constant, $m_1$ and $m_2$ are the masses, and r is the distance of separation. These two force laws are very similar in form. QED theory accounts for Coulomb's law by the photon exchange process. Following the lead from the highly successful QED, the concept of electrical charge exchanging photons is replaced with the idea that gravitational 'mass charge' exchanges gravitons. Hence gravitational mass at a fundamental level is simply the ability to emit or absorb gravitons.

For gravity we have gravitons, instead of photons, which are the force exchange particles of gravity. Like the property of electrical charge, the property called gravitational 'mass-charge' determines the number of exchange gravitons. The larger the mass, the greater the number of elementary particles, and the greater the number of gravitons exchanged. Like electromagnetism, the strength of the gravity force decreases with the inverse square of the distance.

This conceptual framework for quantum gravity has been around for some time now, but how are we to merge these simple ideas in a framework that is also compatible with the framework of general relativity? We must be able to explain the Einstein's Principle of Equivalence and the physical connection between inertia, gravity, and curved 4D space-time, and all within the general framework of graviton particle exchange. General relativity is based on the idea that the forces experienced in a gravitational field and the forces due to acceleration are equivalent, and both are due to 4D space-time curvature.

In classical electromagnetism, if a charged particle is accelerated towards an opposite charged particle, the rate of acceleration depends on the electrical charge value. If the charge is doubled, the force doubles, and the rate of acceleration is doubled. If quantum gravity were to work in the exact same way, we would expect that the rate of acceleration of a mass near the earth would double if the mass doubles. The reason for this expectation is that the exchange process for gravitons should be very similar to electromagnetism. In other words, if the 'mass-charge' is doubled, the gravitational force is doubled. The only difference between the two forces is that gravity is a lot weaker by a factor of about $10^{-40}$ (when comparing the relative strength of the electric and gravitational force for two electrons interacting both electrically and gravitationally).

The weakness of the gravitational forces might be attributed either to the very small interaction cross-section of the graviton particle as compared to the photon particle, or to a very weak coupling constant (the absorption of a single graviton causing a minute amount of acceleration), or for both of these reasons. All test masses accelerate at the same rate (g=9.8 m/sec$^2$ on the earth) no matter what the value of the test mass is. This is a direct consequence of the principle of equivalence. Mathematically, this follows from Newton's two ***different*** force laws: inertial force and gravitational force as follows:

$F_i = m_i a_i$ ..... (Inertial force) (11.21)
$F_g = G m_g M_e / r^2$ ..... (Gravitational force) (11.22)

In free fall, an object of mass $m_i$ in the presence of the earth's pull (Earth's mass $M_e$) is force free, and therefore $F_i = F_g$ (this is also known experimentally). Note that according to Newton, the mass value appears in these *two very* different mass definition formulas for some *mysterious* reason.

Therefore, $ma = GmM_e / r^2$, and since the inertial acceleration is chosen $a = GM_e / r^2$, therefore $m_i = m_g$, ..... the Mass Equivalence Principle.

From this we see that the rate of acceleration does not depend on the value of the test mass m. All test masses accelerate at the same rate on the earth. Thus inertia and gravity are intimately connected in a deep way because the measure of mass m is the same for accelerated frames, as it is for gravity.

What is mass? Recall that according to EMQG, gravitational mass originates from a low-level graviton exchange process due to the property 'mass charge'. In fact, mass is quantized in exactly the same way as electric charge is in QED. Furthermore there exists a fundamental unit of 'mass charge' that is carried by the masseon particle, which is the lowest possible quanta of 'mass charge'.

We have already found an explanation for inertia that is based on low-level quantum processes (section 6). The quantum source of the force of inertia is the resistance to acceleration offered by the virtual electrically charged particles of the quantum vacuum. What is unique about EMQG theory, is that this ***same*** electrical virtual particle force process found in Newtonian Inertia is also present for a mass subjected to a large gravitational field. It is the interactions of these electrically charged virtual particles with the real electrically charged particles in the mass that accounts for almost all of the gravitational force present for that mass. This idea ties in with mass equivalence, thus showing that electrical forces are very prominent in gravitational processes on the earth on the quantum level. Yet, we still retained the same simple QED model for the fundamental low-level gravitational interactions through the graviton exchange process. In fact, we found that the graviton now takes on quantum numbers that are almost the same as the photon. Let us apply this idea to understanding Newton's mass equivalence principle in detail.

## 12. EQUIVALENCE OF INERTIAL MASS AND GRAVITATIONAL MASS

How does the quantum vacuum look near the earth? In other words, is it in the same state of motion as in outer space, away from all masses? The answer to this question is no. On the earth, the virtual particles of the quantum vacuum falls or accelerates downwards during their very brief lifetimes. This is because the virtual mass (masseon) particles are all accelerating towards the center of the earth (at **a**=GM/**r**$^2$) due to direct graviton exchanges between the huge numbers of real mass particles (masseons) that make up the earth and the virtual mass particles (masseons) of the quantum vacuum. During their brief lifetimes, the electrically charged virtual mass particles of the quantum vacuum are capable of interacting with any test mass that is subjected to this. This means that a test mass at rest on the earth will be subjected to this accelerated motion of the quantum vacuum, and therefore the mass ought to 'feel' some kind of inertial force.

In fact, the physics of the vacuum force is the *same* as for accelerated masses in quantum inertia, but now with the acceleration frame of the virtual charged masseons and the real charged masseon particles of the mass being interchanged (with the minor exception of the direct graviton exchanges between the mass and the earth, which is negligible compared to the vacuum electrical forces, see note and end of this section). Equivalence between the inertial mass 'M' on a rocket moving with acceleration 'A', and gravitational mass 'M' under the influence of a gravitational field with acceleration 'A' can now be seen to follow from Newton's laws of inertia and gravitation by slightly rearranging the force formulas as follows:

$F_i$ = M(A)  inertial force which opposes the acceleration A of the mass 'M' in rocket.
$F_g$ = M(GM$_e$/r$^2$) gravity force where ***GM$_e$/r$^2$ now represents virtual particle acceleration***.

Under gravity, the magnitude of the acceleration of the virtual particles subjected to earth's gravitational field is: A=GM$_e$/r$^2$, which is the same as the magnitude of the acceleration chosen for the acceleration of the rocket carrying mass 'M'. Now everything is crystal clear. For a mass 'M' on the floor of the accelerating rocket, each electrically charged mass particle making up the mass interacts with the electrically charged virtual particles of the quantum vacuum resulting in an inertial force. For a stationary mass 'M' on the surface of the earth, each electrically charged mass particle making up the mass interacts with the electrically charged virtual particles of the quantum vacuum (now accelerating downwards with respect to the mass) resulting in a gravitational force that is virtually identical to inertial force.

Another way to look at this is to note that the state of acceleration of the quantum vacuum appears the same for a mass subjected to an accelerated frame, and for a gravitational frame. Equivalence holds because GM$_e$/r$^2$ represents the net statistical average downward acceleration vector of the virtual masseons with respect to the mass 'M', and is **equal** to the acceleration of mass in the rocket. Newton's law of gravity was rearranged here to emphasize the form F=MA for *gravitational mass,* so that we can see that the **same** electrical force summation process for a mass subjected to a gravitational field.

*Note:* **There is a slight force imbalance in perfect mass equivalence due to large numbers of graviton particles, that originate from the earth's mass, being directly absorbed by the stationary mass on the earth (which is not present for the same mass on the accelerated rocket This violation of perfect equivalence is a testable consequence of EMQG theory (ref. 1). This added graviton imbalance is extremely small for the following reason: The electrically charged vacuum particles exert electrical forces similar to inertial force, which tends to drag each electrically charged mass particle in the mass at the same rate of downward acceleration. The gravitons also exert an additional force on each particle of the mass, trying to upset the equalized fall rate. Recalling the ratio of electrical to gravitational is on the order of $10^{40}$, the electrical force completely dominates. However, an experiment can be designed to amplify the graviton effect, and possibly lead to a measurement of this effect. The following describes such a thought experiment to measure the imbalance:**

**Imagine two masses; one mass $M_1$ being very large in value, and the other mass $M_2$ is very small ($M_1 \gg M_2$). These two masses are dropped simultaneously in a uniform gravitational field of 1g from a height 'h', and the same pair of masses are also dropped inside a rocket accelerating at 1g, at the same height 'h'. EMQG predicts that there should be a minute deviation in arrival times on the surface of the earth (only) for the two masses, with the heavier mass arriving just slightly ahead of the smaller mass. Of course the rate of fall of two masses in the rocket is exactly the same, since the floor of the rocket comes up to meet the two masses!**

**This is due to a small deviation in the magnitude of the force of gravity on the mass pair (in favor of $M_1$) on the order of $(N_1-N_2)i * \delta$, where $(N_1-N_2)$ is the difference in the low level mass specified in terms of the difference in the number of masseon particles in the two masses, times the single masseon mass 'i', and $\delta$ is the ratio of the gravitational to electromagnetic forces for a single (charged) masseon. This experiment is very difficult to perform on the earth, because $\delta$ is extremely small ($\approx 10^{-40}$), and $\Delta N = (N_1-N_2)$ cannot in practice be made sufficiently large to produce a measurable effect. However, inside the accelerated rocket, the arrival times are <u>exactly</u> identical for the same pair of masses. Again this imbalance is extremely small, because of the dominance of the strong electromagnetic force which is also acting on the electrically charged masseon particles of the two masses from the electrically charged virtual particles of the quantum vacuum. These electrical forces act to stabilize the fall rate, giving us near perfect mass equivalence.**

## 13.   CONCLUSIONS

We introduced a new paradigm for physical reality, which together with quantum field theory restores a great unity to all physics. We have argued that our universe is a vast Cellular Automaton (CA) simulation, the most massively parallel computer model known. All physical phenomena including space, time, matter, and forces are the result of the interactions of *numerical information* patterns, governed by the mathematical laws and the connectivity of the CA. Because of the way the CA functions, all the known global laws of the physics must somehow result from the local mathematical law or program that governs each cell. Each cell of the CA contains this same mathematical law.

In quantum field theory all forces result from the exchange of boson particles. The particle exchange paradigm fits naturally within CA theory, where the boson exchange process represents the transfer of boson information patterns between quantum particles, which are themselves also information patterns. All forces, with gravity being no exception, originate from particle exchange processes. The *photon* is responsible for the electric and magnetic forces, ***and for the inertial force***, and the graviton (along with the photon) is responsible for gravitational force.

We have investigated the hidden quantum processes that are responsible for Newton's laws of motion, and Newton's universal law of gravity by applying Electro-Magnetic Quantum Gravity, a quantum gravity theory that is manifestly compatible with Cellular Automata theory, and also based on a theory of inertia proposed by R. Haisch, A. Rueda, and H. Puthoff, which we modified and called Quantum Inertia. Newton's $2^{nd}$ law of inertia (F=MA) results from the strictly local electrical force interactions of matter, which consists of electrically charged quantum particles, with the surrounding electrically charged virtual particles of the quantum vacuum. The sum of all the tiny electrical forces originating from each charged particle of the mass interacting with the surrounding quantum vacuum, is the source of the <u>total inertial force</u> that opposes accelerated motion in Newton's law: F = MA. Therefore, to accelerate any mass, a force is required to overcome the electrical opposition force from the quantum vacuum. Similarly, Newton's other two laws of motion were derived within this framework.

We found that Quantum Inertia solves the problems and paradoxes of accelerated motion introduced in Mach's principle by suggesting that the state of acceleration of the charged virtual particles of the quantum vacuum with respect to a mass, serves the function of Newton's ***absolute space*** for accelerated masses *only*. For masses moving at constant speeds, the quantum vacuum does not contribute anything that is observable.

Newton's Universal Gravitational Law ($F = Gm_1m_2/r^2$) results from the countless numbers of graviton particles exchanged between two point masses, where the electrically charged virtual particles of the quantum vacuum also plays a *major role* in the magnitude of the gravitational force. Gravity originates from a fundamental property of all mass particles, the 'mass charge' of the particle. Newton's $F = Gm_1m_2/r^2$ for two point masses is caused by continual exchanges of gravitons which are readily emitted and absorbed by particles with 'mass charge', in analogy with the electrical forces and photon exchanges.

Each exchanged graviton imparts a tiny impulse of momentum to an absorbing mass, which after countless exchanges produces a smooth force. Each mass is <u>both</u> an emitter and a receiver of gravitons, because all mass particles possesses the property of 'mass-charge'. The flux of gravitons emitted by a given mass particle is a fixed quantity, that depends only on the magnitude of the charge, and is defined as the *'mass charge'* of the mass particle. Thus, if the charge doubles, the force doubles because twice as many gravitons are exchanged during the force interaction. This two way force interaction process causes the affected mass particles to accelerate towards each other.

The strength of the gravitational force varies as the inverse square of the distance of separation between the two mass particles due to geometric spreading of the graviton particles in the following way: Each mass particle sends and receives gravitons. As a source of gravitons, the mass particle emits gravitons in all directions. The number of gravitons per unit area received at the destination particle distance r decreases by a factor $1/4\pi r^2$ (the surface area of a sphere, since gravitons spread equally in all directions) at a distance 'r'. If the distance doubles, the number of gravitons exchanged decreases by a factor of four, and this is why there is an inverse square law. If the mass $m_1$ (or $m_2$) of a particle is doubled, the 'mass charge' doubles and twice as many gravitons are exchanged. Therefore this gives twice the force. This accounts for the linear product of mass terms in the numerator of the inverse square law. The constant G is the proportionality constant, and can be derived (in principle) if one knows the detailed low level characteristics of the interaction.

The equivalence of inertial and gravitational mass that first originated in Newton's framework, (and later refined by Einstein in his principle of equivalence) results from the reversal of the relative acceleration vectors of the electrically charged matter particles that make up an accelerated mass, <u>with respect</u> to the (net statistical) average acceleration of the electrically charged virtual particles of the quantum vacuum that occurs when one changes from an accelerated frame to a gravitational frame.

Under gravity, the magnitude of the gravitational field acceleration (the virtual mass particle acceleration) is $A=GM_e/r^2$, which is the same as the magnitude of the acceleration chosen for an equivalent accelerating rocket. From the reference frame of an average accelerated virtual mass particle falling on the earth, a virtual particle 'sees' the real particles that constitutes a stationary mass 'M' on the earth accelerating in exactly the same way as an average stationary virtual mass particle in the rocket 'sees' the accelerated particles constituting a mass 'M' on the floor of the rocket. In other words, the virtual particle vacuum state appears the same in both reference frames. Therefore the electrical interaction between the electrically charged particles that make up a mass with the surrounding electrically charged virtual particles of the quantum vacuum (in a state of relative acceleration) results in the same value of gravitational force as it did for the inertial force. Hence we have equivalence of inertial and gravitational mass. What is unique about EMQG theory is that two bosons are responsible for gravity simultaneously, i.e. the graviton and the photon particles.

However, Newton's mass equivalence is ***not perfect!*** There is a direct graviton exchange process between the earth and the stationary mass in the gravitational reference frame, this tends to add a minute amount of additional force component to the mass in the gravitational field, that is not present for the same mass in an accelerated frame. This slight discrepancy in mass equivalence is a direct and testable consequence of EMQG theory.